\documentclass[aps,pra,reprint,twocolumn,superscriptaddress,nofootinbib,floatfix,showpacs]{revtex4-1}
\usepackage{epsfig}
\usepackage{amsmath,amssymb,amsfonts}
\usepackage{hyperref}
\usepackage{mathrsfs}
\usepackage{bbm}
\usepackage{slashed}
\usepackage{graphicx}
\usepackage{verbatim}
\usepackage{url}
\usepackage[normalem]{ulem}

\usepackage[usenames]{color}

\newcommand{\be}{\begin{equation}}
\newcommand{\ee}{\end{equation}}

\newcommand{\bi}{\begin{itemize}}
\newcommand{\ei}{\end{itemize}}
\newcommand{\bea}{\begin{eqnarray}}
\newcommand{\eea}{\end{eqnarray}}
\newcommand{\ud}{\mathrm{d}}

\newcommand{\Ecr}{E_{\textrm{cr}}}
\newcommand{\sinc}{\text{sinc}}

\definecolor{darkgreen}{rgb}{0.1,0.7,0}
\definecolor{bensblue}{rgb}{0.1,0.1,0.9}

\usepackage[T1]{fontenc} \usepackage[latin1]{inputenc}

\begin{document}
\title{Testing numerical implementations of strong field electrodynamics}

\author{C.~N.~Harvey}
\email[]{christopher.harvey@chalmers.se}
\affiliation{Centre for Plasma Physics, Department of Physics and Astronomy, Queen's University Belfast, BT7 1NN, UK}
\affiliation{Department of Applied Physics, Chalmers University of Technology, SE-41296 Gothenburg, Sweden}
\author{A.~Ilderton}
\email[]{anton.ilderton@chalmers.se}
\affiliation{Department of Applied Physics, Chalmers University of Technology, SE-41296 Gothenburg, Sweden}
\author{B.~King}
\email[]{ben.king@plymouth.ac.uk}
\affiliation{School of Computing and Mathematics, Plymouth University,  Plymouth PL4 8AA, United Kingdom\\}

\begin{abstract}
We test current numerical implementations of laser-matter interactions by comparison with exact analytical results. Focussing on photon emission processes, it is found that the numerics accurately reproduce analytical emission spectra in all considered regimes, except for the harmonic structures often singled out as the most significant high intensity/multi-photon effects. We find that this discrepancy originates in the use of the locally constant field approximation.
\end{abstract}
\maketitle

\section{Introduction}
The study of classical and quantum electrodynamical (QED) processes in strong background fields is currently a highly active research area, as advances in technology now allow fundamental physics to be tested using intense laser light~\cite{marklund06,Dunne:2008kc,EKK2009,Heinzl:2011ur,Redondo:2010dp,Jaeckel:2010ni,DiPiazza:2012RevModPhys}.

The basic QED processes in strong backgrounds were thoroughly investigated soon after the invention of the laser itself~\cite{Nikishov:1964zza,Nikishov:1964zz}, using simple laser field models. However, there is a limit to the field complexity and the number of reacting particles which can be considered before analytical calculations become too unwieldy. Numerical investigation offers one route out of this impasse.

An increasingly common numerical model of QED in intense fields is based on particle-in-cell (PIC) codes. There are a number of such codes in use~\cite{elmis,osiris,psc}, and while they differ in the details, they share a common implementation of laser-matter interactions; particles in an intense field are propagated classically thorough discrete time steps, and statistical event generators are used to determine, at each step, the likelihood and result of various QED processes~\cite{Bell:2008zzb,Sokolov:2010am,nerush10,blackburn14,mironov14}; for reviews see~\cite{Elkina:2010up,Ridgers2014273}.

In this paper we will attempt to assess the accuracy of these numerical techniques in the high-intensity regime appropriate to modern laser systems which reach, and will soon exceed~\cite{Vulcan,ELI,XCELS}, focal intensities of $10^{22}\,$W/cm$^2$~\cite{Yanovsky:2008}. For some previous comparisons of theory and simulation, see \cite{Elkina:2010up,king13a}.

While existing analytical results are largely confined to scattering probabilities (i.e.\ asymptotic results) in simple field configurations and for small numbers of initial and final state particles, the aim of numerical codes is to simulate, in real time, many-particle interactions in focussed, structured laser pulses. Despite these differences, we will show that it is indeed possible to make a direct comparison between numerical predictions and exact analytical results, and thereby test the assumptions that go into the numerical model. Our interest here is not in providing phenomenological results for comparison with experiment, but in addressing the question of whether the currently-employed statistical modifications to PIC simulations can reproduce well-known and well-understood intensity effects in QED. We can therefore employ a simple beam model in order to have a reliable analytical prediction. Furthermore, we restrict our attention to multiple photon emissions from an electron in an intense field, and consider the best-known signatures of intensity and multi-photon effects from the literature. 
 
The paper is organised as follows. We begin in Sect.~\ref{SECT:REVIEW} by briefly reviewing the main features of the photon emission spectrum in nonlinear Compton scattering. There follows a review of commonly used numerical schemes. In Sect.~\ref{SECT:COMP} we describe the observables of interest and compare the analytical and numerical calculations of these for a variety of laser and electron parameters. An analysis of some differences is given in Sect.~\ref{SECT:CLASS} and we conclude in Sect.~\ref{SECT:CONCS}.

\section{Approaches to strong field QED}\label{SECT:REVIEW}
We consider an electron moving in an electromagnetic field, and its emission of photons. We will model the background as a plane wave, a decision motivated not by phenomenological interest but rather by a desire to perform as rigorous a comparison as possible; everything we need can be calculated exactly in plane waves.

The restriction to photon emission implies neglecting e.g.\ pair production from emitted photons~\cite{nikishov.jetp.1967,Erber:1966vv,Baier:2009it,Heinzl:2010vg,Nousch:2012xe} or pair production via trident~\cite{Ritus:1972nf,baier72,hu10,ilderton11,king13b}. Here it is useful to recall the two standard parameters important in this analysis, $a_{0}= e E/m\omega$ (for electron mass and charge $m$ and $e$, external field strength and frequency $E$ and $\omega$, where $\hbar=c=1$) which quantifies the classical nonlinearity, and $\chi_{e}=e|F\cdot p|/m^{3}\sim \gamma E/\Ecr$, where $\Ecr=m^2/e$~\cite{QEDcriticalfield1,QEDcriticalfield2,QEDcriticalfield3}, which quantifies the quantum nonlinearity~\cite{Ritus:1985}. For $\chi_e\gtrsim1$ quantum effects generally become probable and photon-seeded pair production is more likely to occur. 
One finds (using the constant crossed field model) that electron-seeded pair creation via a virtual photon can be neglected when $a_{0}\gg1$~\cite{baier72, king13b}. Further, radiation reaction effects become important only when $\alpha a_{0}\chi_{e} N_c \approx 1$, for $N_c$ the number of cycles in the field~\cite{Harvey:2011dp,DiPiazza:2012RevModPhys}.

Thus, in order to study photon emissions without additional effects, one should be in a regime where $\chi_e$ is less than unity. We require however $a_0\gg 1$ so that we have strong nonlinear effects, and $\gamma$ may still be large, provided $\chi_e\lesssim 1$. We will therefore restrict ourselves to this regime. We stress though that our analysis can be extended, as will become clear, both into the quantum regime, and to e.g.\ stimulated pair production, which exhibits an emission spectrum similar to that of photon emission \cite{nikishov.jetp.1967,Erber:1966vv,Baier:2009it,Heinzl:2010vg,Nousch:2012xe}. 
 
\subsection{Nonlinear Compton scattering}
With ``nonlinear Compton scattering'', we refer to the emission of a photon from an electron in a plane wave background. For simplicity of calculation and presentation we consider a circularly-polarised, monochromatic plane wave travelling in the positive $z$--direction. The electromagnetic fields are therefore comprised of photons of four-momentum $k_\mu$ where $k\cdot x = \omega (t-z)$, and take the form, writing $\phi=k\cdot x$ from here on,
\be\label{FIELD-DEF}\begin{split}
	{\bf E}_{\text{mono.}}(x) &= E \big(\cos \phi , \,\sin \phi , 0\big) \;, \\
	{\bf B}_{\text{mono.}}(x) &= E \big(-\sin \phi , \,\cos \phi , 0\big) \;.
\end{split}
\ee
The intensity parameter $a_0= eE /m\omega$ characterises the strength of interaction between the wave and electrons.

Consider an electron with initial momentum $p_\mu$ and final momentum $p'_\mu$ after emitting a photon of momentum $k'_\mu$. Due to the periodicity and infinite duration of the wave, the probability~$\mathbb{P}$ of emission is infinite. However, dividing out this infinite factor, the emission rate~$W$ per unit $\phi$ (lightfront time) is finite and can be written as a sum over partial rates $W_n$~\cite{Nikishov:1964zza,Nikishov:1964zz},
\be
	W = \sum_{n=1}^\infty W_n \;, \label{eqn:W}
\ee
where $W_n$ describes emissions in a finite kinematic range that correspond to a ``harmonic''. (Comprehensive discussions can be found in any of~\cite{Nikishov:1964zza,Nikishov:1964zz,Harvey:2009,Boca:2009zz,EKK2009,Heinzl:2009nd,Mackenroth:2010jr,DiPiazza:2012RevModPhys}; we review here only the relevant details.) The total and partial rates can be expressed as integrals over differential rates in the outgoing photon frequency and momentum. Although it is common to plot the spectrum as a function of frequency, a variable arising naturally in the QED calculation is the `lightcone momentum fraction' $\text{x}$,
\be\label{x-def}
	{\text{x}}=\frac{k\cdot k^{\prime}}{k\cdot p^{\prime}} =  \frac{k\cdot k^{\prime}}{k\cdot p - k\cdot k^{\prime}} \;.
\ee
Defining the effective mass squared $m_*^2=m^2(1+a_0^2)$~\cite{Volkov:1935,Sengupta:1949}, it can be shown that the kinematically allowed range of the $n^\text{th}$ harmonic is, see  e.g.~\cite{Harvey:2009}, 
\be\label{y-def}
	0 \leq \text{x} \leq y_n \;, \qquad y_n \equiv n \frac{2k\cdot p}{m_{\ast}^2} \;.
\ee
The $n^\text{th}$ harmonic rate $W_n$ vanishes outside of this range. The harmonic decomposition (\ref{eqn:W}) is due to the periodicity of the beam, but in fact any stretch of uniformly periodic field leads to such effects and provides the experimental signature of e.g.\  the intensity-dependent effective mass~\cite{Harvey:2012mass_shift}. Further, harmonic generation has been searched for and observed in several experiments~\cite{PhysRevA.28.1539,Chen:Nature,Iinuma2005255}.

Note that a given {\it frequency} is not necessarily found in a single harmonic range, as these can overlap or be disjoint in frequency space depending on initial conditions and harmonic number~\cite{Harvey:2009}. The allowed scattered photon frequencies $\omega'_n$ in the $n^\text{th}$ harmonic obey, assuming a head-on collision between the electron and laser to illustrate,
\begin{eqnarray}\label{Johnny-5}
\omega_{n}^{\prime}=\frac{n\omega}{1+j_n (1-\cos{\theta})},\label{nuprime}
\end{eqnarray}
where $\theta$ is the photon scattering angle relative to the electron direction and
\begin{eqnarray}
j_n=\frac{n\omega/m -\gamma\beta +a_0^2\gamma (1-\beta )/2}{\gamma (1+\beta )}.\label{jn}
\end{eqnarray}
The behaviour of $j_n$ characterises much of the spectrum. When $j_n <0$ ($j_n>0$) the emission frequency is maximal for back-scattering, $\theta =\pi$  (forward scattering, $\theta=0$). From (\ref{nuprime}) we see that $\omega_n^{\prime}(\theta=0)=n\omega$, so that for $j_n<0$ the scattered frequencies $\omega'_n$ are blue shifted relative to $n\,\times\,$the laser frequency, and if $j_n>0$ they are red shifted.  

The emission spectrum acquires some particularly distinctive features for initial conditions such that $j_n =0$ for some $n$. In that case $\omega^{\prime}_n$ loses its $\theta$-dependence and the $n^\text{th}$ harmonic range collapses to a single point $\omega'_n=n\omega$. Neighbouring harmonics also collapse to very narrow peaks, resulting in a line-spectrum region (a $\delta$-comb like structure) within the full spectrum. Setting $j_n=0$ in order for the $n^\text{th}$ harmonic to collapse, we find that $a_0$ and $\gamma$ must be related by
\begin{eqnarray}
	a_{\textrm{0,crit}}^2(n)\equiv\frac{2(\gamma\beta -n\omega/m)}{\gamma (1-\beta )}.\label{a0crit}
\end{eqnarray}
The behaviour of $j_n$ gives us a useful separation of the system into three parameter regimes: `sub-critical', $a_0<a_{\textrm{0,crit}}$; `super-critical', $a_0>a_{\textrm{0,crit}}$ and `critical', $a_0\approx a_{\textrm{0,crit}}$. These three cases will be examined numerically, below.

\subsection{The numerical approach}
Two limitations on what can be achieved using exact analytical methods are as follows. The first is that the external fields in which scattering amplitudes can currently be calculated exactly do not include the spatial focussing of laser fields employed in experiments.  
The second limitation is that the complexity of $S$-matrix elements grows rapidly with the number of initial and final particles. A complete analytical description of a cascade~\cite{Bell:2008zzb,Sokolov:2010am,nerush10,blackburn14,mironov14}, for example, is extremely challenging. We therefore turn now to the numerical models which offer a route to potentially overcome these limitations. We will outline the shared general principles of currently employed codes, reviewed in~\cite{Elkina:2010up,Ridgers2014273}, beginning with the use of the locally-constant-field (LCF) approximation.

In the high-intensity limit $a_0\gg 1$ the size of the radiation formation region is of the order $\lambda/a_0\ll\lambda$, where $\lambda=2\pi/\omega$ is the laser wavelength~\cite{Ritus:1985}.  Thus the laser field varies on a scale much larger than the formation region and so can be approximated as locally constant and crossed~\cite{Nikishov:1964zza}, allowing us to determine the probability of photon emission using the expression for the constant crossed field rate $\Gamma$ per unit time,
\begin{eqnarray}\label{eqn:dGam}
	\frac{\ud \Gamma}{\ud\chi_\gamma}=\frac{\alpha m}{\sqrt{3}\pi\gamma\chi_e}
\bigg[\bigg( &2&+\frac{\text{x}^2}{1+\text{x}}\bigg) K_{2/3}(\tilde{\chi}) \nonumber\\
\label{constantfieldrate} && -\int_{\tilde{\chi}}^\infty \ud y\, K_{1/3}(y)\bigg],
\end{eqnarray}
where $K_\nu$ is the modified Bessel function, $\chi_\gamma= e|F\cdot k'|/m^3$ for the emitted photon with momentum $k'_\mu$, note that $\text{x}=\chi_\gamma/(\chi_e-\chi_\gamma)$, and $\tilde{\chi}= 2\text{x} /(3\chi_e)$. Although $\ud\Gamma/\ud\chi_{\gamma}$ diverges at small $\chi_{\gamma}$, the total rate of photon emission $\Gamma$, given by integrating (\ref{eqn:dGam}) over all $\chi_\gamma\in[0,\chi_e]$, is finite. (This apparent softening of the usual infra-red divergence in QED is explained in~\cite{Ilderton:2012qe}.)

In numerical simulations, the electron is evolved along a classical trajectory over discrete time steps. After each step  $\Delta t$ the following statistical routine is used to calculate the probability of photon emission and to correct the electron's momentum. A uniform random number $r\in[0,1]$ is generated, and emission deemed to occur if the condition $ r\leq \Gamma \Delta t$ is satisfied, under the requirement $\Gamma \Delta t\ll1$.  Note that $\ud\Gamma/\ud\chi_{\gamma}$ (and $\Gamma$) are time-dependent quantities in the simulation, due to the temporal variation of the laser pulse and electron motion. Given that an emission event occurs, a second uniform random number $\zeta\in[0,1]$ is generated and the photon's $\chi_{\gamma}$ (and therefore its frequency) is determined as the root of the sampling equation\footnote{In practice an infra-red cut-off is used, i.e.\  the integral is performed from a lower limit $\chi_{\gamma} \sim 10^{-5}$, rather than zero, so that the codes do not include the emission of large numbers of low energy photons, which does not appreciably affect the electron's dynamics~\cite{Elkina:2010up,Duclous}. For an alternative event generator see e.g.~\cite{Elkina:2010up}.}
\be
	\zeta={\Gamma(t)}^{-1} \int_{0}^{\chi_{\gamma}}\!\ud\chi_\gamma \frac{\ud \Gamma(t)}{\ud\chi_\gamma} \;.
\ee
The photon momentum is then determined by $\chi_\gamma$ together with the assumption that the electron emits in the forward direction at high $\gamma$. In reality the emissions will be concentrated in a cone of opening angle $\gamma^{-1}$~\cite{jackson_classical_1999, Harvey:2009}. Finally, the emitted photon momentum is subtracted from the electron momentum, i.e.\  the electron is recoiled, imposing the conservation law $\chi_e\to \chi_e-\chi_{\gamma}$~\cite{Ritus:1985}, and the simulation proceeds by propagating the electron (via the Lorentz equation) and the photon (on a linear trajectory) to the next time step. In this way, multiple emissions are described as sequential single photon emissions, as in (\ref{eqn:dGam}), occurring at discrete time intervals.

\section{Comparison of analytical and numerical results}\label{SECT:COMP}
\begin{figure}[t!]
\includegraphics[width=1.0\columnwidth]{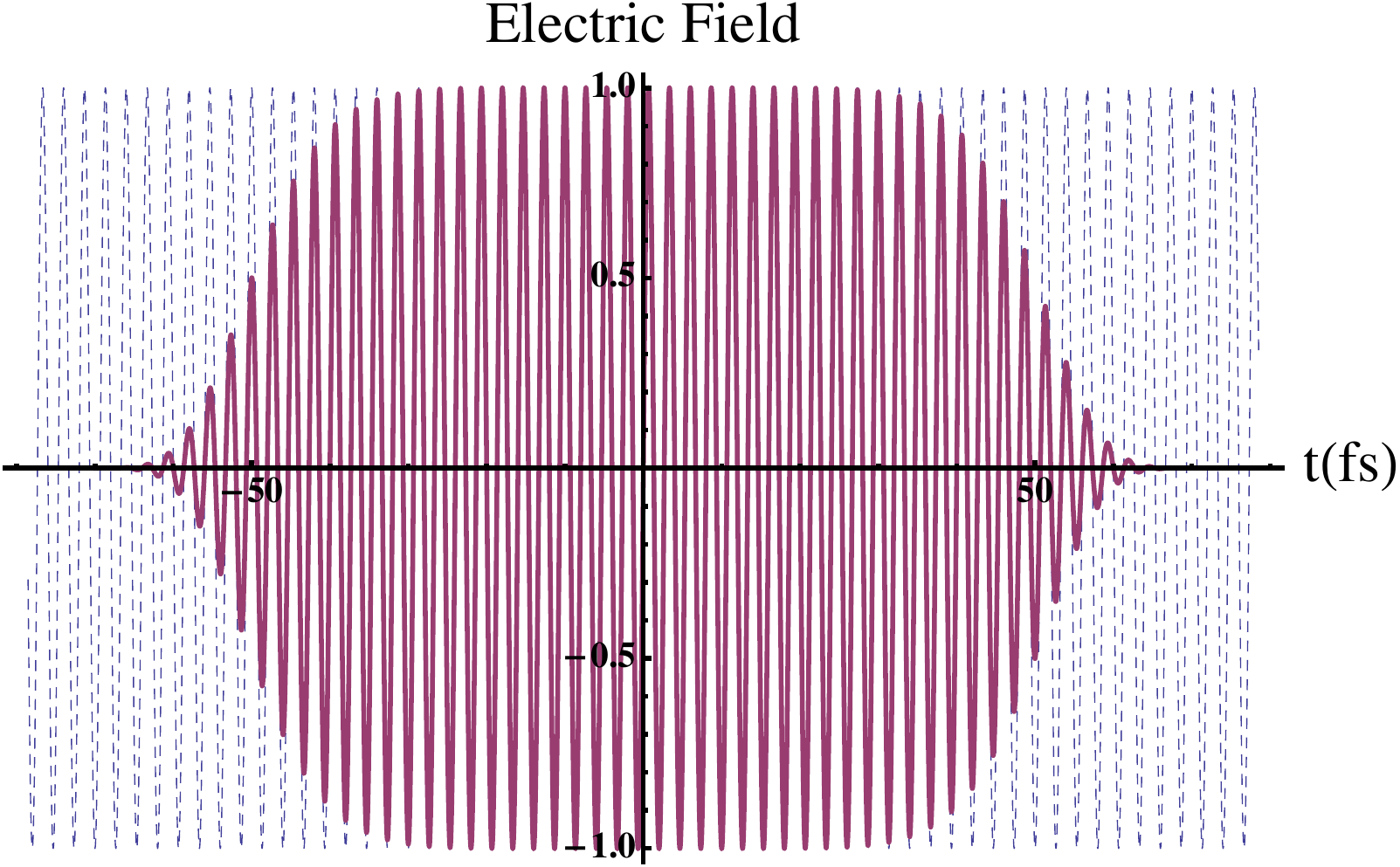}
\caption{Comparison of the electric field strength (normalised to peak field strength) in one of the polarisation directions for an infinite plane wave (dashed/blue line) and a degree 8 super-Gaussian wave of 100\,fs duration (solid/red line) ($\lambda=0.8\mu$m).\label{fig:supergauss} }
\end{figure}

The numerical model described above is not equivalent to the calculation of transition probabilities in QED. In QED, scattering amplitudes are determined using asymptotic in and out states. No assumptions are made about the electron's trajectory in the laser. In the numerical method, on the other hand, we constantly track the electron's trajectory, and asymptotic scattering results are combined with statistical routines to determine the likelihood of local transitions. Therefore, it should be checked whether the numerical model agrees with theory. 

We will calculate and compare analytical and numerical predictions for the observable  $N_\gamma$, the average number of emitted photons; this is clearly something which can be easily extracted from the numerics; we run the code many times, always with the same initial conditions, and count the number of photons emitted into a particular (binned) angle with a particular (binned) frequency.

It may not be immediately obvious how to perform the analytical calculation, as we have described only the single-photon emission probability, and have already mentioned the difficulty with going to higher orders. Fortunately, in the regime in which we are working, knowing the one-photon emission probability is enough because of the well-known infra-red (IR) properties of electrodynamics~\cite{Glauber:1951zz,Yennie:1961ad,Horan:1999ba}, see~\cite[\S 6]{Peskin:1995ev} for an introduction.

The `probability' $\mathbb{P}$ can only be interpreted as such when it is much less than unity, due to neglected higher-order corrections from multiple soft photon emissions\footnote{IR effects in single and multiple photon emissions in background fields are studied in~\cite{DiPiazza:2010mv,Dinu:2012tj,Ilderton:2012qe}.}. It is more properly interpreted as the {\it expected number of emitted photons} $N_\gamma$, which is why it easily exceeds unity~\cite{Peskin:1995ev}. Similarly, the differential probability gives the differential number of produced photons, so
\be\label{PNY}
	\mathbb{P} \to N_\gamma = \int\!\ud\omega\ud\Omega\  \frac{\ud^2 N_\gamma}{\ud\omega \ud\Omega}  \;.
\ee
This allows us to consider multiple emissions using the well-understood probability of single emission. This, and the identification (\ref{PNY}), holds for low energies, as can be confirmed by calculating $N_\gamma$ classically and comparing with the classical limit of the QED emission probability, as in~\cite{Dinu:2013hsd}.

For the purposes of this study we use the single particle QED code SIMLA which works in the typical manner~\cite{Green:2014, SIMLApaper}.  While a monochromatic wave gives the easiest analytical calculation of the emission spectrum, it is more difficult to work with numerically since we cannot run the simulation for an infinite period of time.  However, the spectrum will be very similar to that in a plane wave with a long super-Gaussian time envelope, as demonstrated in~\cite{Harvey:2012mass_shift}.  For our numerical simulations we therefore take a circularly polarised plane wave with a 100\,fs degree-8 super-Gaussian time envelope,
\be
	{\bf E}_\text{super}(x) = e^{-\big(\tfrac{2\phi}{d}\big)^8 \ln{2}}~{\bf E}_\text{mono.}(x) \;,
\ee
in which $d=100\,$fs is the full-width-half-maximum and this is practically equal to the full duration of the pulse, i.e.\ the pulse is almost flat top, as shown in Fig.~\ref{fig:supergauss}.  We now investigate three parameter regimes and present illustrative comparisons between theory and numerics. (In the following plots the analytical amplitudes are normalised using the numerical results.)

\begin{figure}[t!!]
\includegraphics[width=1.0\columnwidth,clip=true,viewport=40 190 540 590]{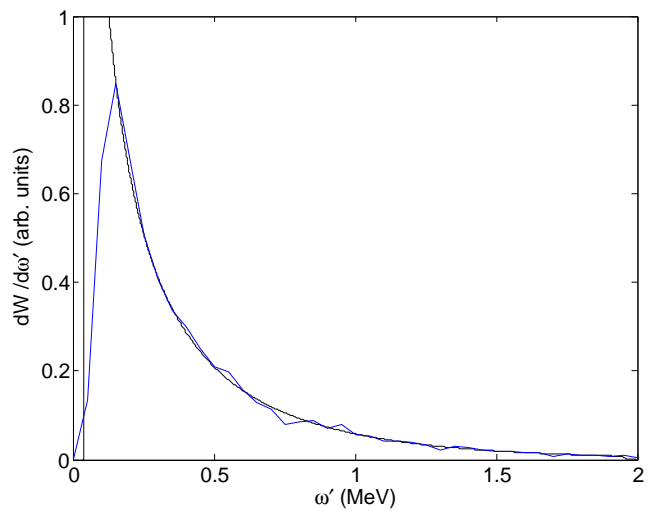}
\caption{Comparison of analytical and numerical frequency emission spectra in the supercritical regime.  The parameters are $a_0=80$, $\gamma=10$. The black line shows the analytical calculation, the blue line the statistical photon distribution from several thousand numerical runs.  \label{fig:a0_80_g_10_freq} }
\end{figure}
\begin{figure}[h!!]
\includegraphics[width=1.0\columnwidth,clip=true,viewport=40 190 540 590]{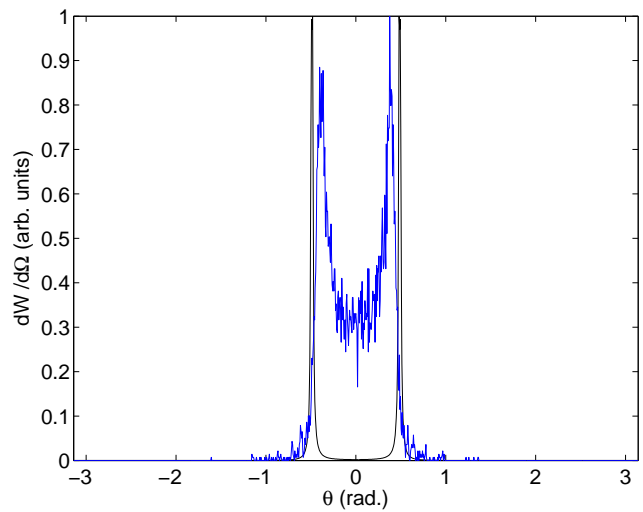}
\caption{Comparison of analytical and numerical angular emission spectra in the supercritical regime. The parameters are $a_0=80$, $\gamma=10$. The black line shows the analytical calculation, the blue line the statistical photon distribution from several thousand numerical runs. \label{fig:a0_80_g_10_angle} }
\end{figure}

\subsection{Supercritical}
We begin with a supercritical setup, $a_0>a_{\textrm{0,crit}}(1)$ (which gives $a_0>a_{\textrm{0,crit}}(n)~\forall\,n \in \mathbb{N}^{+}$). Specifically, we choose the parameters $a_0=80$, $\gamma=10$ such that  $a_0>a_{\textrm{0,crit}}(1)\approx 20$.  The analytical and numerical spectra are plotted in Figs.~\ref{fig:a0_80_g_10_freq} and \ref{fig:a0_80_g_10_angle}.  It can be seen in Fig.~\ref{fig:a0_80_g_10_freq} that the frequency spectra calculated using the two methods agree extremely well.  Both have the same structures and shape and both decay at the same rate.  The only difference is that the numerical spectra falls off for low $\omega^\prime$, but this is simply because of the IR cut-off in the code and should not worry us. Note that the low energy cut-off is not sharp because it is defined in terms of a minimum $\chi_{\gamma}$ rather than frequency\footnote{Additionally, there will be some noise at the low frequency end of the spectrum from when the electron is in the lower intensity rise and fall of the super-Gaussian field.}. 

The angular emission rates are plotted in Fig.~\ref{fig:a0_80_g_10_angle} where it can be seen that there is also fairly good agreement.  The peaks for the numerical and analytical cases are in approximately the same locations and the rate of fall-off for angles larger than the peak value is almost identical.  Nevertheless the two spectra disagree for small angles.  The reason for this is likely to be due to the fact that the code assumes that photons are emitted in the direction of motion of the electron, whereas in reality would be emitted in a cone of opening angle $1/\gamma \sim 0.1$ radians.

\subsection{Subcritical}
\begin{figure}[t!]
\includegraphics[width=1.0\columnwidth,clip=true,viewport=40 190 545 590]{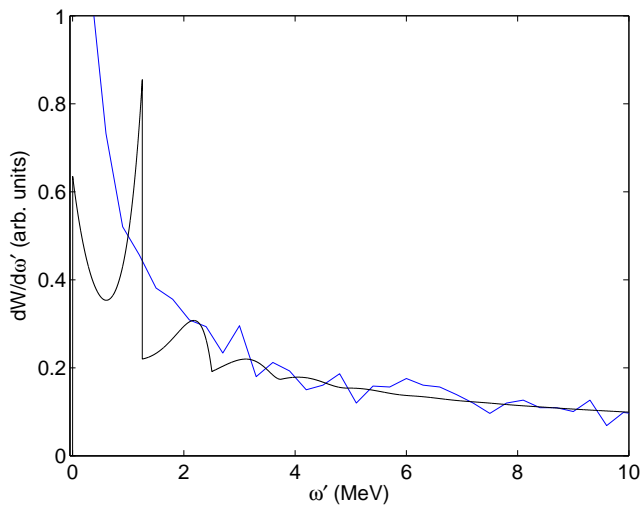}
\caption{Comparison of the analytical and numerical frequency spectra for the case of sub-critical $a_0$.  The parameters are  $a_0=20$, $\gamma=9000$. Black line: analytical spectrum, blue line numerical spectrum. \label{fig:a0_20_g_9000_freq}}
\end{figure}

\begin{figure}
\includegraphics[width=1.0\columnwidth,clip=true,viewport=30 200 545 610]{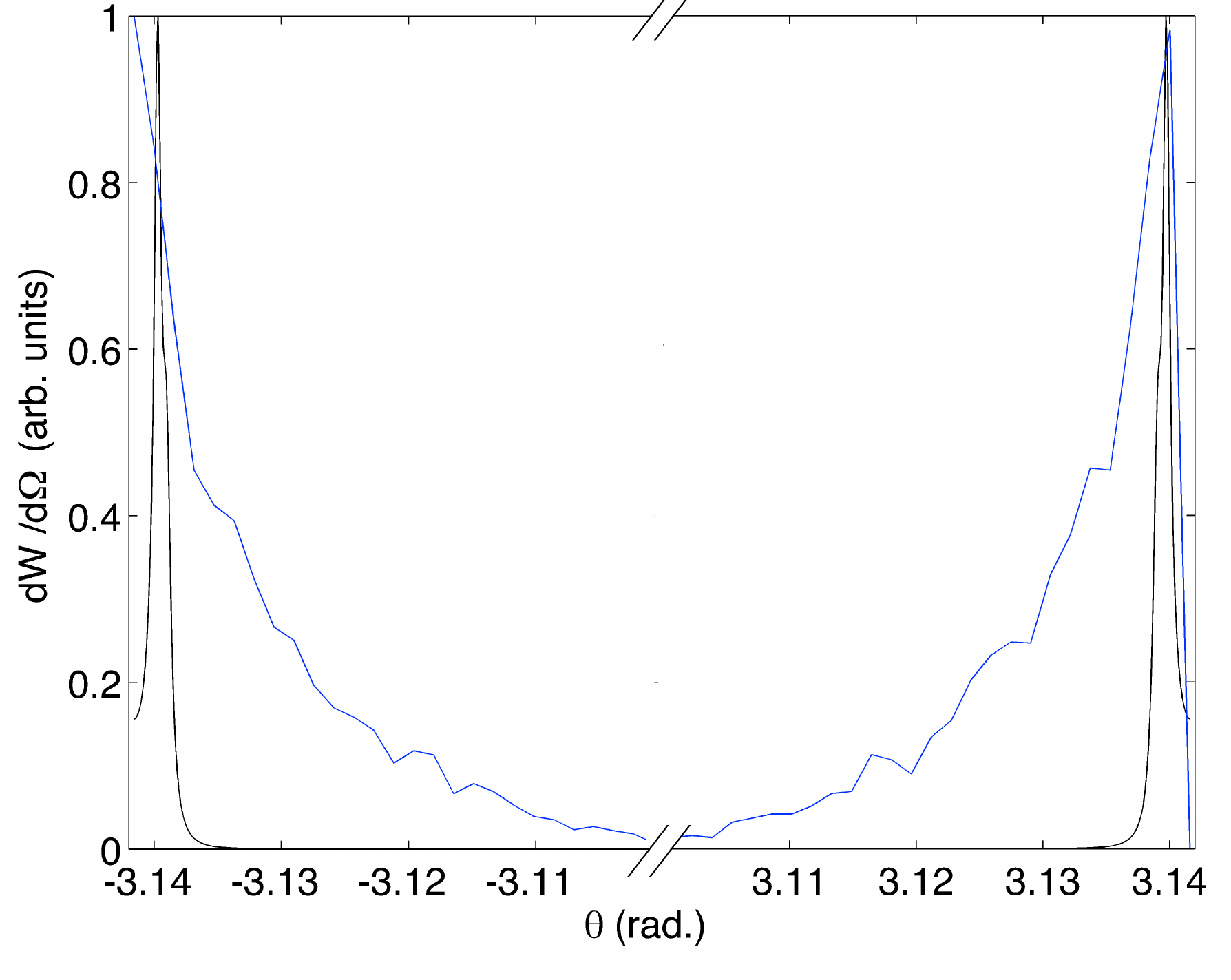}
\caption{\label{fig:a0_20_g_9000_angle} Comparison of the analytical and numerical angular emission rates for the case of sub-critical $a_0$.  The parameters are  $a_0=20$, $\gamma=9000$. Black line: analytical spectrum, blue line numerical spectrum. }
\end{figure}

The next setup we consider is the subcritical regime in which much of the clearly identifiable structure  in the emission spectrum is found in the low-lying harmonics~\cite{Harvey:2009}.  These, being characterised by frequencies equal to multiples of the laser frequency (with an intensity-dependent red-shift, see (\ref{Johnny-5})) are often located below the infra-red cutoff imposed in the codes. In order to lift these into the part of the spectrum resolved by the code we choose $\gamma=9000$ and again take $a_0=20$ ($\chi_e\simeq 0.4$). For these parameters all harmonics up to $n\approx10^{9}$ are blue-shifted; due to the falloff of the spectrum we can effectively say this applies to all harmonics.  The frequency spectra are plotted in Fig.~\ref{fig:a0_20_g_9000_freq} and the angular rates in Fig.~\ref{fig:a0_20_g_9000_angle}.  In both cases there is, in general, a very good agreement between the numerical and analytical results. However we see in Fig.~\ref{fig:a0_20_g_9000_freq} that the structure of individual harmonics, which in this case sit at the lower frequency end of the spectrum, are missed by the code; instead there appears to be a smooth interpolation through them. 

\subsection{Critical}

In the previous example we chose parameters to blue shift the low-lying harmonics to high-frequency. This required very high energy electrons, but even for moderate (readily attainable) energies and laser intensities the emission spectrum can naturally feature interesting structures which should be reproduced by the codes, as they fall into resolved energy regimes. One such feature is described by a collision at `critical' parameters in which a harmonic in the middle of the spectrum collapses.
  
We set the laser intensity to be $a_0=30$ and take $\gamma=15.078$. (There is a degree of fine-tuning here.) From~(\ref{a0crit}) we see that this corresponds to a collapse of the particular harmonic $n=35000$, at $\omega^\prime =0.05415$ MeV. The value of $n$ is irrelevant, what is important is the form of the spectrum; a harmonic collapses to a peak at a single point in the high-energy part of the spectrum, and neighbouring harmonics collapse to very narrow lines, resulting in a region of the spectrum resembling a $\delta$-comb~\cite{Heinzl:2010vg,Krajewska:2014fwa}.

\begin{figure}
\includegraphics[width=1.0\columnwidth,clip=true,viewport=30 190 550 590]{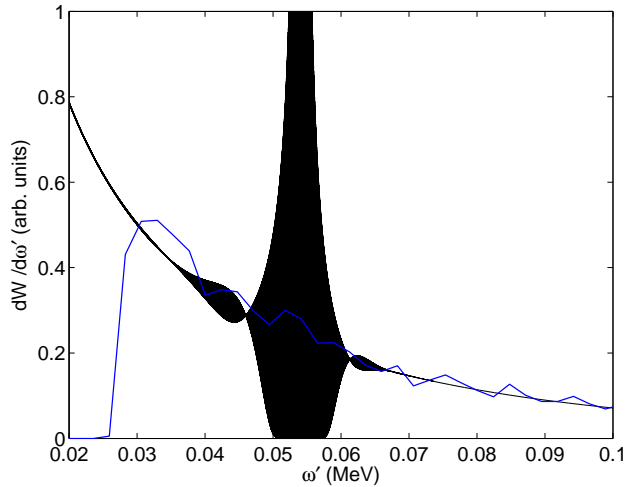}
\caption{Comparison of the analytical and numerical  frequency spectra for the case of critical $a_0$.  The parameters are $a_0=30$, $\gamma=15.078$.  Black line: analytical spectrum, blue line numerical spectrum. \label{fig:a0_30_g_15_078_freq} }
\end{figure}

\begin{figure}
\includegraphics[width=1.0\columnwidth,clip=true,viewport=40 190 560 590]{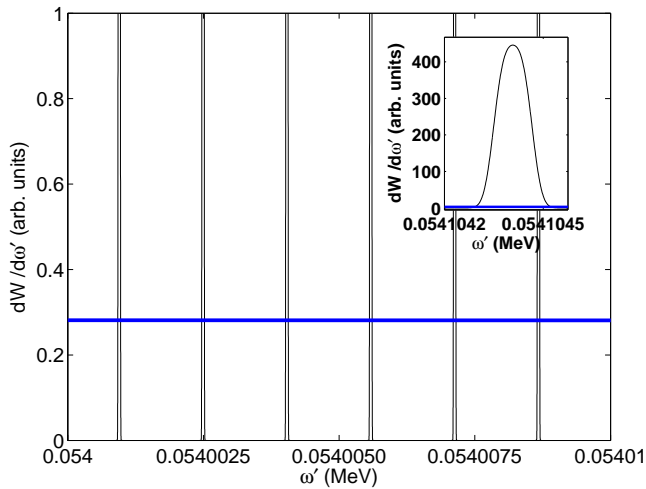}
\caption{A zoom-in of the collapsed part of the spectrum in Fig.~\ref{fig:a0_30_g_15_078_freq}  ($a_0=30$, $\gamma=15.078$).  The inset shows a further zoom into an individual harmonic. \label{fig:a0_30_g_15_078_freq_zoom} }
\end{figure}

\begin{figure}
\includegraphics[width=1.0\columnwidth,clip=true,viewport=40 190 545 590]{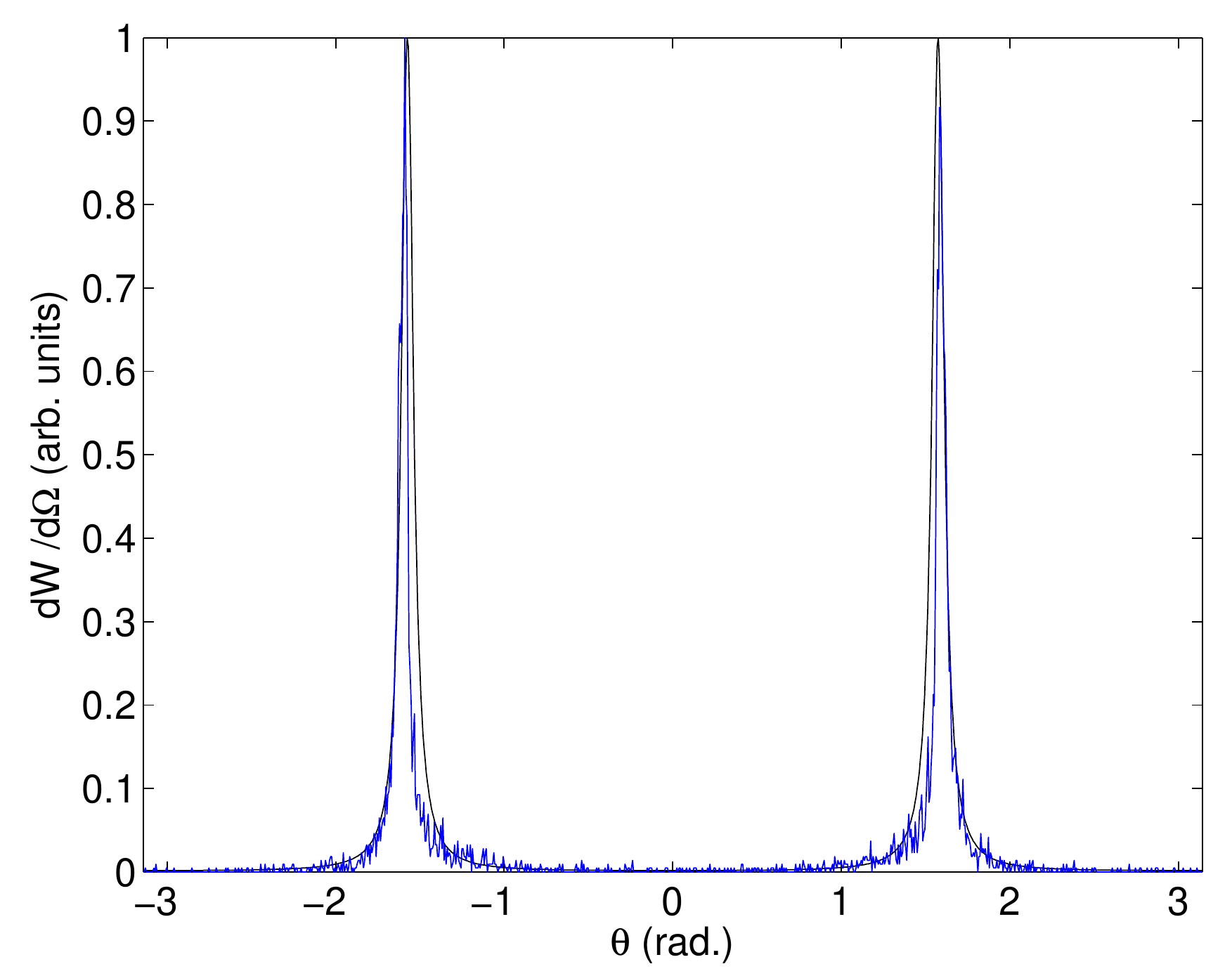}
\caption{Comparison of the analytical and numerical angular emission rates for the case of critical $a_0$.  The parameters are $a_0=30$, $\gamma=15.078$.  Black line: analytical spectrum, blue line numerical spectrum. \label{fig:a0_30_g_15_078_angle}}
\end{figure}

The resulting emission spectra are plotted in Figs.~\ref{fig:a0_30_g_15_078_freq}-\ref{fig:a0_30_g_15_078_angle}.  In Fig.~\ref{fig:a0_30_g_15_078_freq} the peaked and rapidly oscillating comb-like structure can be very clearly seen. Away from this part of the spectrum, the analytical and numerical results agree very well once again. In the region of the comb structure however, and as can be seen clearly in the zoomed-in plot Fig.~\ref{fig:a0_30_g_15_078_freq_zoom}, the numerical spectrum again does not resolve the harmonic peaks in the analytical spectrum; instead it appears to average over them. As can be seen in Fig.~\ref{fig:a0_30_g_15_078_angle}, though, the analytical and numerical angular rates agree extremely well.

\begin{figure}[t!]
\includegraphics[width=1.0\columnwidth,clip=true,viewport=40 190 545 590]{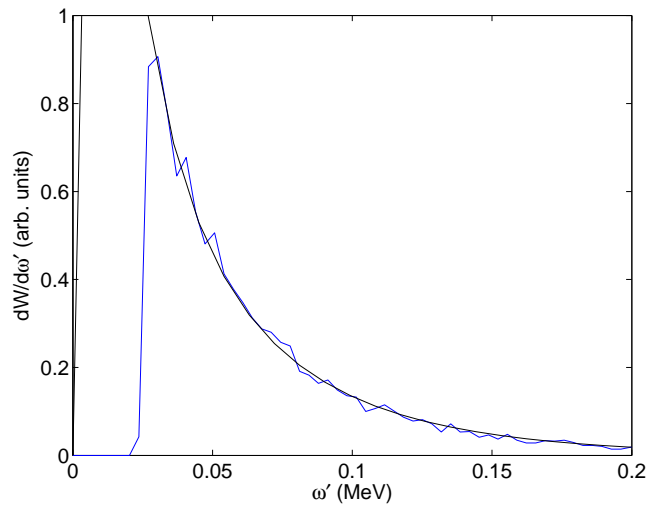}
\caption{\label{fig:a0_30_g_16_freq} Comparison of the analytical and numerical angular emission rates for the case of {\it near}-critical $a_0$. The parameters are $a_0=30$, $\gamma=16$. Black line: analytical spectrum, blue line numerical spectrum. }
\end{figure}

It is instructive to also consider parameters that are slightly ``off critical'' in order to demonstrate how sensitive the spectrum is to initial conditions.  We take $a_0=30$ and $\gamma=16$, so that the comb-like part of the spectrum is shifted just out of the region we are considering. The frequency spectrum is shown in Fig.~\ref{fig:a0_30_g_16_freq}.  Here the individual harmonics join together to produce a continuous spectrum, and we can see that the numerical results once again agree extremely well with the analytical  calculation. (To shift the `critical' harmonic beyond 0.2 MeV in Fig.~\ref{fig:a0_30_g_16_freq} one needs to increase $\gamma$ to 15.123. The spectral range in a plane wave is infinite, but the spectrum decreases exponentially at high frequency.)


\section{Analysis}\label{SECT:CLASS}
Although the numerical model captures the coarse details of the emission spectra considered, it fails to reproduce any of the harmonic signatures that characterise emission in monochromatic waves. 
We have found that this discrepancy is best explained in terms of the LCF approximation. In accordance with earlier discussions we analyse this in the classical limit.

While the extension to quantum expressions will be apparent, we postpone a detailed analysis to future work.

\begin{figure}[t!]
\includegraphics[width=\columnwidth]{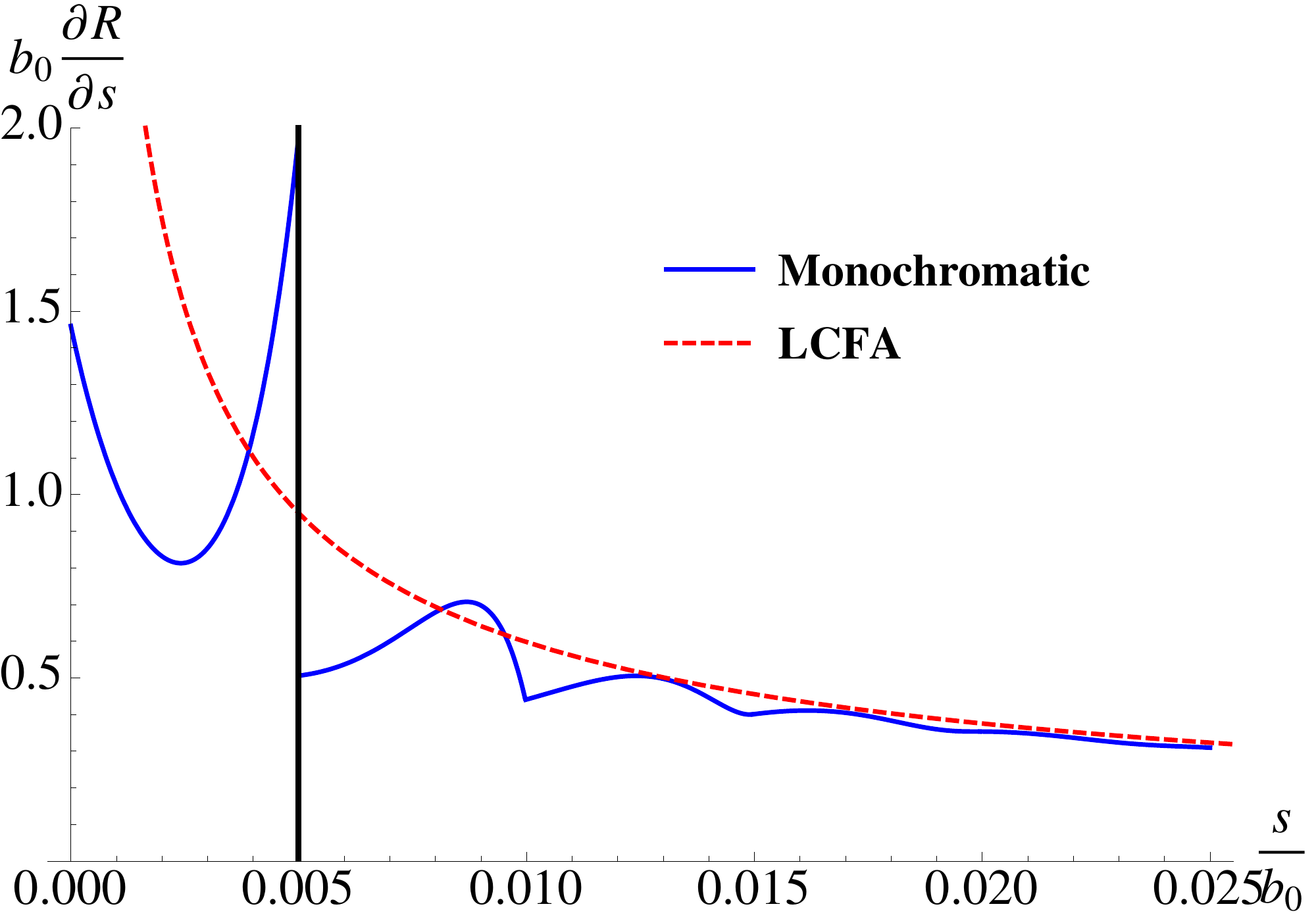}
\caption{\label{FIG:KLASS} The full spectrum for a monochromatic field (blue/solid) and the LCF approximation to it (red/dashed). $a_0=20$ to illustrate. The vertical black line marks the boundary of the first harmonic.}
\end{figure}

The total number of emitted photons, $N_\gamma$, can be calculated either from standard expressions~\cite[\S 14]{jackson_classical_1999}, or from the classical limit of the nonlinear Compton scattering probability calculated in QED, as in~\cite{Dinu:2013hsd}. Since $N_{\gamma}$ is given by an integral over the mod-square electromagnetic current generated by the electron~\cite{jackson_classical_1999}, it can be written as a double integral over the external-field phase $\phi$,
\bea
	N_{\gamma} = \int\! \ud\phi\, \ud\phi'~ \frac{\ud^{2} N_{\gamma}}{\ud\phi\,\ud\phi'} \;. \label{eqn:dNdphi}
\eea
We define $\sigma=\phi+\phi'$ and $\tau = (\phi'-\phi)/2$, respectively double the average of, and half the time between, emissions at different points on the electron trajectory. In a circularly-polarised monochromatic plane wave, the integrand above is independent of $\sigma$ and therefore (as already discussed), infinite. However, the rate $W=N_{\gamma}/\int \ud\sigma$ is finite, which gives
\bea
	W = \int\!\ud \tau~\frac{\ud W}{\ud \tau} \;.
\eea
In order to compare with our numerical results we require the differential rate with respect to a kinematic variable, and for this we choose the low-energy (classical) limit of the variable (\ref{x-def}), which is
\be
	s= \frac{k\cdot k'}{k \cdot p} = \frac{\chi_\gamma}{\chi_e} \;,
\ee
as can be checked by reintroducing $\hbar$ and expanding in powers of the photon momentum. The differential rate (summed over all harmonics) we will consider is
\be\label{DIFF-RATE}\begin{split}
	\frac{\partial W}{\partial s} = -\frac{\alpha}{2\pi b_0} \int_{-\infty}^{\infty}\!\frac{\ud \tau}{\tau} &\bigg(\sin \frac{\tau s}{b_0}-\sin \frac{\tau s\mu}{b_0} \\
	&-2\tau^2a_0^2\,\sinc^2 \tau \,\sin \frac{\tau s\mu}{b_0} \bigg) \;,
\end{split}
\ee
in which $\tau$ is as above, $b_{0}=k\cdot p/m^{2}$  and $\mu$ is the ratio of Kibble's mass $M^2$ to $m^2$~\cite{Kibble:1975vz} in a monochromatic wave being equal to
\be
	M^2(\tau)/m^2 = \mu(\tau) = 1 + a_0^2\big(1 - \text{sinc}^2 \tau\big) \;.
\ee
Note that the asymptotic limit of $M^2$ is the effective mass squared, $m_{\ast}^2$, introduced above.
Since the integrand in (\ref{DIFF-RATE}) is even in $\tau$, we are dealing with integrals of the form 
\be\label{integral-shape}
	\int_{0}^{\infty}\!\ud \tau \ f(\tau) e^{i\Phi(\tau)} \;, 
\ee
where the phase
\be
	\Phi(\tau) = \frac{s\tau}{b_{0}}\big[1+a_{0}^{2}\big(1-\sinc^2 \tau\big)\big] \;,
\ee
has no extrema on the real line. The LCF approximation corresponds to expanding $\Phi(t)$ to next-to-lowest order in~$\tau$, giving
\bea \label{eqn:PhiLCFA}
	\Phi(\tau) \simeq \Phi_\text{LCF}(\tau) = \frac{s\tau}{b_{0}} \left[1+\frac{a_0^2 \tau^2}{3}\right] \;.
\eea
Since the condition for neglecting the next term in this expansion is $ \tau^{2} \ll 15/2 $, we see that using the expansion in~(\ref{eqn:PhiLCFA}) assumes that only a finite range $t\in[0,\tau_{\text{max}}]$ contributes to photon emission and that the total integral can be approximated as
\bea \label{eqn:t_ass}
	\int_{0}^{\infty}\!\ud \tau \ f(\tau) e^{i\Phi(\tau)} \approx 	\int_{0}^{\tau_{\text{max}}} \ud \tau\ f(\tau) e^{i \Phi_\text{LCF}(\tau)}.
\eea
Recall that $\tau$ measures the correlation of radiation emitted from different points on the electron trajectory. If only small $\tau$ is included, then emission is assumed to be ``local''. Writing $x^{-}=t-z$, we note that $\tau=k\cdot(x'-x)/2=\pi(x'^{-}-x^{-})/\lambda$ for external field wavelength $\lambda$, and observe that the small-$\tau$ limit can be understood as the large-$\lambda$ limit, thereby demonstrating the equivalence of the LCF approximation and the constant crossed field limit in the current problem. When the LCF approximation is applied to (\ref{DIFF-RATE}), we recover the constant crossed field expression upon performing the $\tau$-integral,
\be\label{DIFF-LCF}
	\frac{\partial W_{\text{LCF}}}{\partial s} = -\frac{\alpha}{b_0}\bigg( \text{Ai}_1(z) + \frac{2}{z}\text{Ai}'(z)\bigg)\;, 
\ee
where $z=(s/a_0 b_0)^{2/3}$. Now, if the LCF assumptions approximate the original integral well, the majority of the full integral in $t$ must be included in the assumption (\ref{eqn:t_ass}), which implies $s\tau_\text{max}/{b_0} \gtrsim 1$. If $\tau_{\text{max}} \sim O(1)$, this condition can be fulfilled for $s/b_{0} \gg 1$. However, $s$ is not always confined to this range, as is most evident for the first harmonic for which, from (\ref{y-def}), $s/b_{0}\leq 1/(1+a_0^2)<1$. Therefore the LCF assumptions break down here; this is demonstrated in Fig.~\ref{FIG:KLASS}, where the LCF approximation misses the first harmonic structure, precisely as was seen in the numerical simulation Fig.~\ref{fig:a0_20_g_9000_freq}.

In fact we can show explicitly that information pertaining to the first harmonic is contained in the large $\tau$ (i.e.\ {\it large-distance}) expansion of the integrand, in particular in the asymptotic expansion of the effective mass.  To do so we simply replace $M^2$ with its asymptotic limit~$m_\ast^2$ in (\ref{DIFF-RATE}) and evaluate~\cite[\S 3.828-3]{Grad}
\be\begin{split}
	\frac{\partial W}{\partial s} \to& \frac{\alpha a_0^2}{\pi b_0} \int\!\frac{\ud \tau}{\tau}\, \sin^2 \tau ~\sin\frac{2 s\tau}{y_1} \\
	&=\frac{\alpha a_0^2}{2b_0} \, \theta(y_1 -s ) \;,
\end{split}
\ee
(where $\theta(\cdot)$ is the Heaviside function with $\theta(0)=1/2$) which is precisely the range of the first harmonic, and also gives the jump discontinuity clearly visible in the emission spectrum, marked in Fig.~\ref{FIG:KLASS}. The LCF approximation knows nothing about the large distance expansion of the effective mass and is blind to harmonic structure, in particular the first harmonic; the LCF approximation to the emission spectrum, used either analytically as in Fig.~\ref{FIG:KLASS} or numerically as in Fig.~\ref{fig:a0_20_g_9000_freq}, goes smoothly through the jump at the edge of the first harmonic range.

\section{Conclusions}\label{SECT:CONCS}

We have tested current numerical implementations of strong-field electrodynamics by comparing their predicted spectra with known analytical results. Focussing on photon emission from an electron, we have found that common PIC-based models correctly reproduce many features of the emission spectra. The high-energy tails of the distributions are well matched, as are the angular distributions, especially in the highly relativistic limit where the electrons emit almost entirely forward, matching the assumption that goes into the codes.

What the numerical model fails to reproduce is the harmonic structure of the photon distribution in frequency-space. The well-known peaks and troughs associated with (higher) harmonic generation, which are the distinct, indeed measured, signals of intensity and multi-photon effects, are missed. It has been shown that this discrepancy can be attributed to the use of the locally-constant-field approximation, which is the standard way of including strong-field QED processes in numerical models.

Despite the good overall agreement between theory and numerics, our investigation prompts further enquiry. By extending our numerical and analytical results to the quantum regime, other processes such as pair production can also be scrutinised.

\acknowledgments
C.H.~thanks Dermot Green and Adam Noble for useful discussions. C.H.\ is supported by EPSRC, grant EP/I029206/1--YOTTA and A.I.\ by {\it The Swedish Research Council}, contract 2011-4221.

%

\end{document}